\documentstyle[12pt,amsfonts,amssymb,amscd]{report}
\begin{document}
\begin{titlepage}
\pagestyle{empty}
\title{Black Hole Entropy, Topological Entropy and Noncommutative Geometry}
\author{Ioannis P. \ ZOIS\thanks{izois\,@\,maths.ox.ac.uk; Research supported by the EU, contract no HPMF-CT-1999-00088}\\
\\
Mathematical Institute, 24-29 St. Giles', Oxford OX1 3LB\\}
\date{}
\maketitle
\begin{abstract}

\emph{Foliated manifolds} are particular examples of \emph{noncommutative spaces}. In this article we try to give a \emph{qualitative} description of the \emph{Godbillon-Vey class} and its relation on the one hand to the \emph{holonomy} and on the other hand to the \emph{topological entropy} of a foliation, using a remarkable theorem proved recently by G. Duminy relating these three notions in the case of \textsl{codim-1 foliations}. Moreover we shall investigate its possible relation with the \emph{black hole entropy} adopting the superstring theory origin of the black hole entropy in the extremal case. This situation we believe has some striking similarities with the explanation due to Bellissard of the integrality of the Hall conductivity in the quantum Hall effect. Our starting point is the Connes-Douglas-Schwarz article on compactifications of matrix models to noncommutative tori.\\

PACS classification: 11.10.-z; 11.15.-q; 11.30.-Ly\\

Keywords: Godbillon-Vey class, String-Theory, Foliations, Topological Entropy.\\

\end{abstract}
\end{titlepage}

\section{Introduction and motivation}

In a number of papers for some time now (see \cite{Z1}, \cite{Z2}), we have been trying to understand some of the underlying \emph{topology} of M-Theory. Our approach is along the lines of \cite{CDS}, namely \emph{noncommutative geometry}. In the most recent article \cite{Z1} we tried to compute the transition amplitudes between some noncommutative vacua of M-Theory using the Godbillon-Vey class as a Lagrangian density in the particular case of codim-1 foliations of the 3-torus $T^3$. The real motivation for that (from the point of view of physics) was an attempt towards M-\emph{Field} Theory. Our approach uses the notion of \emph{foliated manifolds} which are \textsl{particular examples} of noncommutative spaces. In physics literature the usual framework is that of a noncommutative algebra which is supposed to represent the algebra of the coordinate functions of some noncommutative space. Our foliation approach is not as general but we believe gives more intuition and better understanding. We hope the reader will appreciate that once we shall discuss our application to the Beckenstein-Hawking formula.

In this article, our goal is two-fold: firstly, in order to clarify the motivation behind \cite{Z1}, we shall try to exhibit, mainly via examples, what the role of the Godbillon-Vey class (abreviated to ``GV'' in the sequel) in the geometry of foliated manifolds is and thus give a better physical interpretation for the path integral we tried to calculate in \cite{Z1}. The surprising result is that the \emph{Godbillon-Vey class in a subtle way counts the topological entropy of the foliation}. Secondly, we shall try to see its possible relation with the derivation of the Beckenstein-Hawking area-entropy formula from string theory. ``En route'' we shall attempt to draw some analogies with Chern classes for bundles used to describe the topological charges of instantons in Yang-Mills theory. As it is well known, Chern classes count the ``number of global twists'' of a bundle which represents physically a soliton solution of the Yang-Mills equation and that is related to the energy of the solution (classical vacuum).

\section{Foliations}

Let $M$ be a smooth closed $n$-manifold. A codim-$q$ (and hence dim-$(n-q)$) foliation $F$ on $M$ is given by a codim-$q$ integrable subbundle $F$ of the tangent bundle $TM$ of $M$. Sometimes $F$ itself is called the \emph{tangent bundle of the foliation} and its \textsl{quotient bundle} $\nu (F):=TM/F$ is called the \emph{normal bundle of the foliation}. This defines a \textsl{decomposition} of $M$ into a \textsl{disjoint union} of $(n-q)$-dimensional submanifolds which are called the \emph{leaves} of the foliation. \textsl{Thus since in general the leaves of a foliation do not intersect one can think of a foliation as defining a generalised notion of} \emph{``parallelism''} \textsl{between its leaves}. The simplest and most trivial example of a foliation is of course Cartesian product.\\

The leaves of a foliation have three important properties: (i) they are connected (but may not be of course simply connected) submanifolds of $M$, (ii) they are all of the \textsl{same} dimension $(n-q)$ and (iii) they are \emph{immersed} submanifolds of $M$. One might immediately observe some similarities with the \emph{total space} of a fibre bundle. In fact \textsl{the total space of a fibre bundle is the 2nd simplest example of a foliation}, but in fact still rather a trivial one, the leaves being the fibres. One has two main differences between fibrations and foliations: 1.The fibres are \emph{embedded} submanifolds whereas the leaves need only be \textsl{immersed}; that gives a notion of ``parallelism'' for leaves in a foliation which is far more general than the situation for fibres in a bundle 2.All fibres are usually \emph{diffeomorphic (or homeomorphic)} to some fixed ``model'' manifold which is called the ``typical fibre''. For foliations the situation is \textsl{drastically different}: \emph{the leaves may not have the same fundamental group (!)} hence they may not be even homotopy equivalent and some of them may be compact and some others may not. These two differences will be of paramount importance in our discussion because they will give rise to what is called in foliation theory the \emph{``holonomy''} of the foliation. That in turn is the source of \emph{noncommutativity} on the corresponding $C^*$-algebra of the foliation (see \cite{Z2}). It is worthwile mentioning an analogy here: we would like to think of foliations in noncommutative geometry in some sense as an analogue of symplectic manifolds and Poisson algebras. Any symplectic manifold gives a Poisson algebra structure on its corresponding commutative algebra of functions yet of course not every Poisson algebra can be thought of as coming from a symplectic manifold. One has more Poisson algebras than symplectic manifolds. The use of symplectic geometry though is important because it gives clearer pictures and one can get more insight. Thi
e algebras and foliations. The former are far more general but in studying foliations one uses topology and can get more insight.\\

There is a local definition for a foliated manifold using the notion of \emph{foliated chart and atlas}. A \emph{foliated chart of codim-$q$} on a smooth closed $n$-manifold $M$ is a pair $(U, \phi )$ where $U\subset M$ is open and $\phi :U\rightarrow B_{\tau }\times B_{\pitchfork }$ is a diffeomorphism and $B_{\tau }$ is a rectangular neighborhood in ${\bf R^{n-q}}$ where ``$\tau $'' stands for \textsl{tangential} and $B_{\pitchfork }$ is a rectangular neighborhood in ${\bf R^q}$ where ``$\pitchfork $'' stands for \textsl{transverse}. The set $P_{y}=\phi ^{-1}(B_{\tau }\times \{y\})$ where $y\in B_{\pitchfork }$ is called a \textsl{plaque} of this foliated chart. Similarly for each $x\in B_{\tau }$ the set $S_{x}=\phi ^{-1}(\{x\}\times B_{\pitchfork })$ is called a \textsl{transversal} of the foliated chart. Then a \emph{foliated atlas of codim-$q$} is a collection of foliated charts $\{U_{a}, \phi _{a}\}_{a\in A}$ that cover $M$. \textsl{The plaques patched together form the leaves of the foliation}, hence the leaves have dimension $(n-q)$. If we want to emphasise the transverse or tangential coordinates we shall write $(U_{a}, x_{a}, y_{a})$ for $(U_{a}, \phi _{a})$, with $x_{a}=(x^{1}_{a},...,x^{n-q}_{a})$ i.e. tangential coordinates and $y_{a}=(y^{1}_{a},...,y^{q}_{a})$, i.e. the transverse coordinates. On the overlap $U_{a}\cap U_{b}$ we denote $g_{ab}$ the \emph{tangential} \textsl{transition functions} and $\gamma _{ab}$ the \emph{transverse} \textsl{transition functions}. Only the \emph{transverse} transition functions satisfy the \emph{cocycle conditions}:$$\gamma _{aa}=1$$
$$\gamma _{ab}=\gamma ^{-1}_{ba}$$
and
$$\gamma _{ac}=\gamma _{ab}\circ \gamma _{bc}$$
The set $\gamma =\{\gamma _{ab}\}_{a,b \in A}$ of the \textsl{transverse} transition functions is called the \emph{holonomy cocycle} of the (regular) foliated atlas  $\{U_{a}, \phi _{a}\}_{a\in A}$. (``Regular'' is some technical convenient notion, one can prove that every foliated atlas has a regular refinement).\\

A codim-$q$ foliated $n$-manifold (or a manifold carrying a foliation) is then defined to be an equivalence class of codim-$q$ foliated atlases (after introducing an appropriate notion of equivalence called \emph{coherence}). The proof that this definition using foliated atlases is equivalent to the one mentioned in the beginning involving integrable subbundles of the tangent bundle, is by no means trivial and can be found in for example \cite{CC} (Theorem 1.3.8 Frobenius Theorem p37). Note moreover that this definition of a cocycle is roughly the same used in \cite{Z1} but there we called it a Haefliger or $\Gamma _{q}$-cocycle. There is a difference however, here we assume local diffeomorphisms essentially from ${\bf R^{q}}$ to itself whereas in \cite{Z1} we assumed \textsl{germs} of local diffeomorphisms from ${\bf R^{q}}$ to itself. In the next section where we shall elaborate on the holonomy, we'll see that the first cocycle defines the \textsl{total holonomy pseudogroup} of the foliation whereas the second defines the \textsl{germinal holonomy groupoid} of the foliation; they are of course very closely related and essentially they ``contain'' the same information.\\

The key point to note in this definition is that in some sense one introduces a \emph{topological ``decomposition''} of $M$ to \emph{tangential} and \emph{transverse} directions. Physicists cannot avoid comparing this structure with \emph{supermanifolds} in which case a manifold has also an \emph{algebraic decomposition} into commuting and anticommuting directions (coordinates).\\

 There is a generalisation of the above definition in the case where the \emph{transverse} piece of the manifold is required to be \textsl{homeomorphic to an arbitrary metrizable topological space} (namely not just to some Euclidean space) but still keeping the \textsl{tangential part diffeomorphic to some Euclidean space (hence the leaves will be manifolds)}. In this far more general setting one talks about \emph{foliated spaces} (i.e. not foliated manifolds) or \emph{abstract laminations}. Moreover clearly \textsl{one could think of a supermanifold as a (rather trivial) lamination where the transverse space, corresponding to fermionic degrees of freedom, is a Euclidean space with reversed parity.}\\

 Let us mention also that Frobenius theorem allows one to define a codim-$q$ foliation via a non-singular decomposable $q$-form $\omega $ say, on $M$ which vanishes exactly on vectors tangent to the leaves of the foliation. Integrability then implies that
$$\omega \wedge d\omega =0$$
It is important to underline that \textsl{the leaves of foliations of arbitrary dimension on a manifold $M$ are in fact the higher dimensional generalisations of flows of vector fields of $M$}. Equivalently they can be considered as the \emph{orbits} of a (generalised) dynamical system. Foliations may also have \emph{singularities}. For simplicity we shall not consider this case here.\\

\section{Holonomy}

Let us now elaborate on the notion of the \emph{holonomy} of the foliation. Again, let $M$ be a closed smooth $n$-manifold having a codim-$q$ foliation $F$. What the holonomy of the foliation does is the following: if $L$ is a leaf of the foliation and $s$ a \emph{path} in $L$, but usually we shall consider it to be a loop, one is interested in the behavior of the foliation in a neighborhood of $s$ in $M$. Intuitively we may think of ourselves as ``walking'' along the path $s$ keeping an eye on all the nearby leaves; as we walk we may see some of these nearby leaves ``peeling away'', getting out of visual range, others may suddenly come into range and approach $L$ asymptotically (for simplicity we shall ignore the points of intersection of leaves which would lead to singularities; to deal with singularities one has to use laminations; recall that in general the leaves are not normally allowed to intersect since they are ``parallel''), others may follow along in a more or less parallel fashion or wind around $L$ laterally etc. This behaviour when appropriately formalised is called the \emph{holonomy} of the foliation. There are basically two ways to do that: one is by defining the \emph{total holonomy pseudogroup} of the foliation or by defining the \emph{holonomy groupoid} of the foliation (or what is called \emph{germinal holonomy} in \cite{CC}). The second is due to Wilnkenkempern (see \cite{Wilnkenkempern}) and it is what was used in \cite{Z2} to define the corresponding $C^{*}$-algebra of the foliation and then \emph{a new invariant for foliated manifolds}. One can see all the relevant details in \cite{CC}. The important thing to note here is that \emph{the total holonomy pseudogroup or the holonomy groupoid of the foliation essentially contains all the information coming from the fundamental groups of each one of the leaves plus their configuration}, hence everything concerning the topology of the foliation. The key observation then is that an \textsl{infinitesimal version of the germinal holonomy} is actual
odim-1 case where things are more straightforward. So \emph{the GV-class essentially contains information about the holonomy of the foliation which is in fact responsible for the noncommutativity of the corresponding $C^{*}$-algebra!} \textsl{Yet, as Duminy's theorem states, this is done in a rather complicated way which is yet far from being clearly understood.} \\

In order to understand the notion of the holonomy of a foliation, it is better to give an example; that will involve the famous Reeb foliation (codim-1 case) of the 3-torus (for a formal discussion see \cite{CC} p15 and p45). Take the boundary leaf $L$ of the Reeb foliated solid torus. $L$ itself topologically is a 2-torus. Let $s(t)$ be a longitudinal loop on $L$ of period $a>0$ and based at $p=s(0)$. Imagine yourself walking along $s$. With each complete circuit of $s$, the walker will see the nearby leaves spiral in closer. (The picture for one to have in mind is of ``a snake trying to bite its own tail'' and hence swallowing itself but that happens repeatedly). Our observer could quantify this data by carrying a rod  $J_t$, always \textsl{perpendicular} to the home leaf $L$ and having one endpoint at $s(t)$ (in the general case of a codim-$q$ foliation this will be a \textsl{transversal}; here because we are in codim-1 case it is just an interval). After one circuit $J_{0}=J_a$ and the points of intersection of this rod with nearby leaves will all have moved closer to the endpoint $s(0)=s(a)=p$. This is called ``first return map'' in dynamical systems and can be viewed as a diffeomorphism $h_{s}:J_{0}\rightarrow I$ onto a subinterval $I\subset J_{0}$ which also has $p$ as an end point. This contraction map to $p$ is called the \emph{holonomy} of the loop $s$.\\

In order to give a clearer intuitive picture, let us recall another special example (of arbitrary codimension): perhaps the simplest non-trivial example of a foliation of arbitrary codimension is called a \emph{foliated bundle} and it is in fact a \emph{flat} principal $G$-bundle with total space $M$, structure Lie group $G$ which the fibres are homeomorphic (or diffeomorphic) to and base space $N$ which is \textsl{not simply-connected}. In this case one has \emph{two} structures on the \emph{total space}: the fibration (with fibre $G$) and the foliation induced by the flat connection where the leaves are \emph{covering spaces} $\tilde{N}$ of the base space $N$. This case was studied in great detail in \cite{Z2} where we actually proved that the $C^{*}$-algebras associated to these two structures on the total space were $C(N)\otimes K$ for the fibration and $C(N)\rtimes \pi _1(N)$ for the foliation, where $K$ is the elementary $C^{*}$-algebra of compact operators acting as smoothing kernels along the fibres and $\pi _{1}(N)$ is the fundamental group of the base space $N$. Both these $C^{*}$-algebras are noncommutative but the first is Morita equivalent to just $C(N)$ which of course is commutative whereas the second is \textsl{not} even Morita equivalent to a commutative one. The lesson we would like to extract from this example is the following: \emph{the true origin of the noncommutativity of the corresponding $C^{*}$-algebra of a foliated bundle lies in the fundamental groups of its leaves(!)} which are \textsl{also responsible for the foliation not having trivial total holonomy pseudogroup or trivial holonomy groupoid.} We should keep that in mind for later discussions. (Note: in this example only the fundamental groups play a role in noncommutativity; in full generality ``parallelism'' will also enter the scene).\\

Let us also mention that in the above example of foliated bundles, the total holonomy pseudogroup of the foliation is calculated to be the following: as it is well-known a (gauge equivalence class of a) flat connection corresponds (this is a 1:1 correspondence in fact) to a (conjugacy class of an) irreducible representation $a$ of the fundamental group of the base space to the structure group of the bundle, namely $a:\pi _{1}(N)\rightarrow G$. In this case the total holonomy pseudogroup of the foliated bundle is actually a group, the image of $\pi _{1}(N)$ into $G$ by $a$. Note moreover that in the foliated bundle case the fibration and the foliation are transverse to each other.\\

 An arbitrary foliation of codim-$q$ in general does not admit a global cross section, which by definition is a transversal intersecting all leaves; it certainly admits local ones though. (In the previous example of a foliated bundle all the \textsl{fibres} are \textsl{global cross sections} of the foliation.) Then the above construction can be done only \emph{locally} and not globally; that is the reason why in general we end up with pseudogroups instead of groups and the analogue of $G$ will be $Diff^{+}({\bf R}^{q})$. Going back to our codim-1 example of the Reeb foliation of the solid torus, the holonomy $h_s$ of the path $s(t)$, with $t\in [0,1]$ for a foliation of codim-$q$ will now be a map $h_{s}:S_{s(0)}\rightarrow S_{s(1)}$ where $S_{s(0)}$ and $S_{s(1)}$ are \emph{transversals} (well in fact some open neighborhoods of transversals) of the foliation (the analogues of the ``rod'' $J_0$ and $I$ in the Reeb foliation example; in that case they are 1-dim, hence intervals). All such local diffeomorphisms $h_s$ form a \emph{pseudogroup} denoted $\Gamma _{U}$ relative to the regular foliated atlas $U$. It is a pseudogroup and not a group because transformations are not globally defined, hence composition may not always be defined. (Note however that pseudogroups are closed under the operation of \textsl{amalgamation}). It is clear we think that \textsl{elements of $\Gamma _{U}$} can be thought of as \textsl{transformations between transverse $q$-manifolds} defined by \emph{``sliding alond leaves''}. In fact this is exactly the holonomy cocycle of the foliation. Notice the dependence on the foliated atlas; in fact this is not too bad, essentially all regular atlases of a foliation contain ``the same information''. For a detailed discussion see \cite{H}.\\

The above discusion can be done using the \emph{germs} $\hat{h_s}$ instead of local diffeomorphisms $h_s$. Recall that the germ $\hat{h_s}$ is the equivalence class of all local diffeomorphisms agreeing with $h_s$ in small neighborhoods. Let $S:=\coprod _{a\in A}S_{a}$ be the disjoint union of all the transversals of our codim-$q$ foliation. Then if $y$, $z$ are in $S$ and they lie on the same leaf we define $G_{y}^{z}:=\{h_{s}|s$ a path in $L$ from $y$ to $z$\}. If they do not lie on a common leaf we set $G_{y}^{z}:=0$. One has a \emph{natural composition} $G_{y}^{z}\times G_{x}^{y}\rightarrow G_{x}^{z}$ defined by $(\hat{h}, \hat{f})\rightarrow \hat{h\circ f}$. Moreover for each $x\in S$ $G_{x}^{x}$ contains the identity. Thus $G_{S}:=\coprod _{y,z \in S}G_{y}^{z}$ is a \emph{groupoid} (a small category with inverses) and it is called the \emph{holonomy groupoid} or the  \emph{germinal holonomy} of the foliation. $G_S$ is in fact an element of the set $H^{1}(M;\Gamma _{q})$ according to the terminology of \cite{Z1} and by dividing this by a homotopy equivalence relation one gets the topological category denoted $\Gamma _{q}(M)$ in \cite{Z1}. Then $\Gamma _{q}(-)$ is a homotopy invariant functor and it was used to define K-Theory No2 in \cite{Z1}, where $M$ is our smooth closed $n$-manifold carrying the codim-$q$ foliations. If $y\in S$ and $L$ is the leaf of our foliation $F$ through $y$, then $G_{y}^{y}$ is actually a \emph{group} and it is called the \emph{holonomy group} of the leaf $L$ at $y$ and it will be denoted $H_{y}(L)$. If $s$ is a loop on $L$ based at $y\in L\cap S$, then $\hat{h_{s}}\in H_{y}(L)$. This defines a surjective map which is in fact a group homomorphism (for the proof see \cite{CC} p60)
$$\hat{h}:\pi (L,y)\rightarrow H_{y}(L)$$
This is called the \emph{germinal holonomy of the leaf $L$}.\\

Let us close this section by remarking that another way to think of foliations is as spaces carrying a $G$-action but now $G$ is not a Lie group as it is the case in Yang-Mills theory but it is actually a \textsl{pseudogroup} or a \textsl{groupoid} (the ones attached to the foliation using its holonomy). So they can be thought of as generalising principal $G$-bundles for $G$ being a groupoid. This approach has been used in \cite{Connes}. Finally another important point to remember is that foliations from the mathematical point of view are not the most general structures arising from bundles but they are the most general structures known up to now that one can still do K-Theory, define characteristic classes (using the Gelfand-Fuchs cohomology) and eventually have index theorems for leafwise elliptic operators.\\

{\bf Aside:} It was argued in \cite{SW} that Morita equivalent algebras give rise to the same physics. This combined with the fact that Morita equivalent algebras have the same K-Theory and cyclic cohomology actually suggest that from the noncommutative topology point of view an algebra should not only be noncommutative but more than that it should not even be Morita equivalent to a commutative one in order to be interesting from the noncommutative topology point of view. Also recall that from Serre-Swan theorem the K-Theory of commutative $C^{*}$-algebras coincides with Atiyah's original K-Theory for spaces and moreover cyclic cohomology for commutative algebras (almost) coincides with the de Rham cohomology of the underlying space.\\

Summarizing then, in this section we tried to exhibit mainly via examples, that the noncommutativity in foliations arises from the fact that the leaves of a foliation may have different fundamental groups and the notion of the holonomy of the foliation essentially captures the information of the fundamental groups of all of its leaves. Moreover one has a more general notion of parallelism since the leaves are only immersed submanifolds and not (as is the case for fibrations) embedded.

\section{The Godbillon-Vey class}

The purpose of this section is to try to give a flavour of ``what the Godbillon-Vey class does'' for foliations geometricaly. We know for instance that \textsl{the Chern classes for bundles essentially count the number of twists} that the bundle may have and thus give a non-trivial topology. This fact is used in Yang-Mills equations to ``tell one soliton solution from another'', where the Chern classes are in fact the topological charges for the soliton solutions. Yet this topological charge is directly related to the \emph{energy}, hence between topologically distinct vacua there is an energy barrier, roughly analogous to the \emph{difference} of the topological charges (in fact 2nd Chern classes) of the two vacua (bundles) involved; using the quantum mechanical property of barrier penetration then one can give a physical interpretation of instantons. In \cite{Z1} we tried to compute some \emph{transition amplitudes} between \textsl{noncommutative vacua} using the GV-class as a Lagrangian density. We knew that these noncommutative vacua can be seen as \textsl{topologically distinct foliations} of the underlying manifold, hence the \textsl{GV-class} can indeed be used as a \emph{topological charge} to distinguish one solution from another. Its physical interpretation though was not clear. Here we try to understand the GV-class better in geometric terms and then discuss its possible physical interpetation.

So the story goes as follows: for convenience, let us restrict ourselves to the \emph{codim-1} case where things are clearer. A foliation $F$ on a smooth $n$-manifold $M$ as in the previous section can be defined by a non-singular 1-form $\omega $ vanishing exactly at vectors tangent to the leaves. Integrability of the corresponding $(n-1)$-plane bundle $F$ of $TM$ implies that $\omega \wedge d\omega =0$ or equivalently $d\omega =\omega \wedge \eta $ where $\eta $ is another 1-form. The 3-form $\eta \wedge d\eta $ is closed hence determines a de Rham cohomology class called the \emph{Godbillon-Vey} class of $F$. Although $\omega $ is only determined by $F$ up to multiplication by nowhere vanishing functions and $\eta $ is determined by $\omega $ only up to adition of a $d$-exact form, actually the Godbillon-Vey class depends only on the foliation $F$. The Godbillon-Vey class can also be defined for foliations of codim grater than 1 as we explained in \cite{Z1} and equivalently $\eta $ can be thought of as a \emph{basic} (or sometimes called \emph{Bott}) connection on the normal bundle.\\

Now here is the {\bf key} \emph{observation:} for a codim-$q$ foliation $F$, the \textsl{Jacobian} $J$ of the \emph{germinal holonomy homomorphism $h$ of a leaf} $L$ of $F$ defines a group homomorphism $Jh:\pi _{1}(L)\rightarrow GL(q;{\bf R})$ called \emph{infinitesimal germinal holonomy homomorphism} of the leaf $L$. For \emph{codim-1 case} one gets simply ${\bf R}^{+}$ as the range of the group homomorphism and it is customary in this case to assume composition with $log$ function and eventually get a map from the fundamental group of the leaf to ${\bf R}$. On the other hand, in algebraic topology there is a standard identification of the first cohomology group $H^{1}(L;{\bf R})$ with the set of group homomorphisms $\psi :\pi _{1}(L)\rightarrow {\bf R}$ so one can view the composition $log\circ Jh$ as an element of $H^{1}(L;{\bf R})$. In particular when $H^{1}(L;{\bf R})=0$, the infinitesimal germinal holonomy of the leaf is trivial. The 1-form $\eta $ used in the definition of the GV-class restricted to $L$ is \emph{closed} and its class in $H^{1}(L;{\bf R})$ is exactly the infinitesimal germinal holonomy class $log\circ Jh$! This suggests that \emph{$GV(F)=[\eta \wedge d\eta ]\in H^{3}(M;{\bf R})$ should be tied with the holonomy of the foliation!}  ({\bf Aside:} \emph{This observation largely explains the similarities observed in \cite{Z1} between codim-1 foliations of a 3-manifold defined by \textsl{closed 1-forms} and \textsl{Abelian Chern-Simons} theory} on-shell; ``on-shell'' in Chern-Simons theory means ``flat'' and hence in the Abelian case just ``closed''). At this point we meet Duminy's theorem saying that this is indeed true but the relation between the GV-class and holonomy is not very straightforward.\\

{\bf Note:} In general the infinitesimal germinal holonomy of a leaf $L$ of a codim-$q$ foliation $F$ on $M$ is an element of $Hom(\pi _{1}(L);GL(q;{\bf R}))$. For codim-1 case then triviality of infinitesimal holonomy is equivalent to vanishing of just $H^{1}(L;{\bf R})$. This key observation is partly responsible for very special things occuring in codim-1 case, one of the most impressive ones being the famous Thurston stability theorem (which improves Reeb stability).\\

Before stating and explaining Duminy's result let us recall some facts about the \textsl{delicate} issue of the \emph{invariance} of the GV-class. Recall that by definition a foliation is an \textsl{integrable} subbundle of the tangent bundle. Hence roughly one could define \textsl{two notions of homotopy} equivalence between foliations: the first assumes integrability in the intermediate steps whereas the second neglects it. To be more precise, there are actually \textsl{three} useful equivalence relations between foliations (starting from the most general and going to the most narrow these are): homotopy, concordance and integrable homotopy.\\

The precise definitions can be found in \cite{CC}. Here we shall try to give an intuitive picture. We shall say that two foliations $F$ and $F'$ of our manifold $M$ are \emph{integrable homotopic} if one can deform continously one to the other \textsl{through intermediate foliations $F_t$} where $t\in [0,1]$ with $F_{0}=F$ and $F_{1}=F'$. There is a variant of this definition using the notion of \emph{isotopy} but for compact $M$ they coincide. If we divide the set $H^{1}(M;\Gamma _{q})$ by integrable homotopy equivalence relation, we end up with the topological category denoted $Fol_{q}(M)$ in \cite{Bott}. This then can in principle be used to define another K-Theory using the Quillen-Segal construction as described in \cite{Z1}.\\

The second definition is the following:  we shall say that two foliations $F$ and $F'$ of our manifold $M$ are \emph{concordant} if one can deform continously one to the other through \emph{Haefliger structures} $F_t$ where $t\in [0,1]$ with $F_{0}=F$ and $F_{1}=F'$. A Haefliger structure is a mild generalisation of a foliation. We shall call the foliations simply \emph{homotopic} if the deformation is performed through arbitrary (i.e. not necessarily integrable) subbundles of the tangent bundle. \emph{This} homotopy takes us from $H^{1}(M;\Gamma _{q})$ to $\Gamma _{q}(M)$ according to the terminology in \cite{Z1}.\\

Now it is a surprising (perhaps) fact that although the GV-class is a de Rham cohomology class (and de Rham cohomology is homotopy invariant), it is not homotopy invariant according to the notion of homotopy defined just above for foliations. The GV-class is only \emph{integrable homotopy} invariant. In the particular case of \textsl{codim-1 foliations} it is \emph{concordance} invariant. If our manifold $M$ is a \textsl{compact 3-manifold}, then the \emph{GV-invariant} (i.e. the real number obtained from evaluation of GV-class against the fundamental 3-homology class $[M]$ of $M$), is in fact \emph{cobordism} invariant (more general than concordance). For the proofs the interested reader may see \cite{CC} (Chapter 3).\\

Moreover the GV-class may vary continously and non-trivially; in fact Thurston has proved in \cite{th} that for the 3-sphere the GV-invariant can take any non-negative real value! (again for a simplified version of this proof see \cite{CC}). This result is probably not very encouraging for the invariant we tried to define in \cite{Z1}; the point here though is that in \cite{Z1} we restricted ourselves to \emph{taut} codim-1 foliations (for reasons coming both from physics and from mathematics) and $S^{3}$ has \emph{no taut codim-1 foliations} essentially because all codim-1 foliations of the 3-sphere have a Reeb component. We would like to note that foliations in general are rather ``too flexible'' structures and one usually wants to make them more ``rigid'' and one way to do that is to restrict to ``taut'' ones as described in \cite{Z1}. Just recall that on a closed oriented connected smooth 3-manifold $M$, a codim-1 foliation is called (geometrically) taut if $M$ admits a Riemannian metric for which all leaves are minimal surfaces or equivalently if there exists a closed transversal (that cannot be anything else than $S^{1}$) which intersects all leaves. There is yet another characterisation of (geometrically) taut foliations due to Rummler in this case using forms, namely for each taut codim-1 foliation there exists a unique closed 2-form which is transverse to the foliation (namely it is non-singular when restricted to the leaves of the foliation). There is an analogous statement which goes under the name of Sullivan's theorem for codim $>1$ taut foliations which we shall mention in the Appendix.\\

One last definition before we state Duminy's theorem: a leaf $L$ of a codim-1 foliation $F$ is called \emph{resilient} if there exists a transverse arc $J=[x,y)$ where $x\in L$ and a loop $s$ on $L$ based on $x$ such that $h_{s}:[x,y)\rightarrow [x,y)$ is a contraction to $x$ and the intersection of $L$ and $(x,y)$ is non-empty.  (Note that in the definition above the arc $J$ is \textsl{transverse} to the foliation). Intuitively a resilient leaf is one that ``captures itself by a holonomy contraction''. The terminology comes from the French word \emph{``ressort''} which means \emph{``spring-like''}.\\

Now we shall state {\bf Duminy's Theorem:}(see \cite{D})\\

\textsl{For a codim-1 foliation $F$ on a closed smooth $n$-manifold $M$ one has that $GV(F)=0$ unless $F$ has some (at least one)} \emph{resilient} leaves.\\

The proof of this theorem is still unpublished. A discussion about the proof has been exhibited in \cite{CC} using the theory of \emph{levels (or ``architecture'') of foliations}. The authors mentioned in \cite{CC} that the full proof of Duminy's theorem {\bf will appear} in the forthcoming Vol II of their book. We give a brief outline of the proof in the Appendix. The point here is that although in the previous section we mentioned the observation that the GV-class is related to the infinitesimal holonomy of the leaves of the foliation, now Duminy's theorem makes a far more delicate statement saying that it is in fact \emph{only the resilient leaves} that \textsl{contribute to the GV-class} of the foliation!\\

The notion of resilient leaves can be generalised for foliations with codim grater than 1 although we do not know what their relation with the GV-class would be in these cases.\\

Let us treat some special cases as examples: The GV-class \emph{vanishes} for a special class of codim-1 foliations, those defined by \emph{closed} 1-forms. ({\bf Remark:} this is true for foliations defined by \emph{constant} differential forms which is the case considered in the Connes-Douglas-Schwarz article \cite{CDS}). We suspect that it is probably also zero in general for \emph{fibrations} although we have not been able to prove this. We mentioned that a codim-1 foliation $F$ on a closed smooth $n$-manifold $M$ can be defined by a non-singular 1-form $\omega $ on $M$. Integrability implies $\omega \wedge d\omega =0$. In particular, this is automatically satisfied for \emph{closed} 1-forms. Now codim-1 foliations defined by closed 1-forms have a very special property: they are \emph{homeomorphic} to foliations with \textsl{no holonomy}, see for example \cite{CC} section 9.3. It is rather obvious, as was also pointed out in \cite{Z1} when comparing codim-1 foliations on a 3-manifold and Abelian Chern-Simons theory, that \textsl{a codim-1 foliation defined by a closed 1-form} (i.e. Abelian Chern-Simons on-shell) \textsl{has vanishing GV-class}. Now if the codim-1 foliation has trivial holonomy then it is homeomorphic to a foliation defined by a closed 1-form but homeomorphisms do not respect smooth forms and hence we cannot conclude that the GV-class vanishes. This is \textsl{true however} being a corrolary to Duminy's theorem since such foliations \textsl{cannot have resilient leaves} (see \cite{morita}).\\

The relation that all this has with bundles is the following: a codim-1 foliation $F$ defined by a closed 1-form $\omega $ on $M$ has automatically a \emph{transverse holonomy invariant measure} $\mu $, where $\omega :=d\mu $ is closed. Denote by $P(\mu )$ the group of \emph{periods} of the measure $\mu $ which by definition is the image of the group homomorphism $[\mu ]:\pi _{1}(M)\rightarrow {\bf R}$ (and hence itself can be seen also as an element of $H^{1}(M;{\bf R})$) defined by
$$[\mu ]([\sigma ]):=\int _{\sigma }d\mu $$

One can prove that as cohomology classes $[\omega ]=[\mu ]$ in $H^{1}(M;{\bf R})$ (see \cite{CC} Proposition 9.3.4 p219).\\

The \emph{rank} $p(\mu )$ of $\mu $ is then by definition the rank of the group $P(\mu )$. Then one has\\

{\bf Proposition 1:}\\
$F$ is a fibration if and only if $p(\mu )=1$.\\

 (For the proof see \cite{CC} p219). Hence one can get a characterisation of $M$ being a codim-1 foliated manifold or the total space of a fibration. The later is a special case of a codim-1 foliation defined by a closed 1-form. Moreover a theorem of Tischler states that codim-1 foliations defined by closed non-singular 1-forms can be smoothly and uniformly ``well approximated'' by fibrations over $S^{1}$ (see \cite{CC} p221). Note moreover that if a closed smooth $n$-manifold admits a codim-1 foliation defined by a closed 1-form that would imply that its first cohomology group does not vanish.\\

The lesson from all this discussion is that at least in the codim-1 case, foliations defined by \emph{closed} 1-forms are \textsl{``the next closest thing to a fibre bundle''}, namely they are homeomorphic to foliations with no holonomy and have vanishing GV-class. In dimension 3 we know that the classification of manifolds is not category dependent hence there is no difference between homeomorphism and diffeomorphism and consequently in dim 3 a foliation with no holonomy is the same as a foliation defined by a closed 1-form. Recall that in the previous section we mentioned that \emph{the GV-class measures in a subtle way the noncommutativity of the foliation} (whose origin is the holonomy); here \textsl{we saw an indication of the validity of that statement} (at least for the codim-1 case), coming up in a complicated way: in \cite{Z2} we saw that the corresponding $C^{*}$-algebra of an \textsl{arbitrary} fibre bundle, although it is noncommutative, it is in fact \emph{Morita equivalent to a commutative one} (hence, seen as a foliation, it has no holonomy); on the other hand, codim-1 foliations defined by \textsl{closed 1-forms are very closely related (in fact they are homeomorphic) to codim-1 foliations having no holonomy}, a \textsl{special case} of which are \textsl{fibrations over $S^{1}$} and finally \textsl{in all these cases the GV-class} \emph{vanishes} indeed (as a consequence of Duminy's theorem).\\

\section{Topological Entropy}

Our discussion above concerned primarily codim-1 foliations. This will still be mainly the case in this section also. Let us start with the following remark: in higher codimensions things become much more complicated and the study of foliations involves \emph{statistical techniques} borrowed from \emph{ergodic theory}. It is in this framework that the important notion of \emph{topological entropy} arose historically for the first time. The fundamental result in this section is the remarkable consequence of Duminy's theorem (recall that this holds for codim-1 foliations) that: \emph{``non-vanishing GV-class implies positive entropy''} for the foliation!\\

To begin with we shall define the notion of \emph{entropy of maps} and then we shall generalise it for foliations using as intermediate steps the entropy of transformation groups and pseudogroups.\\

 In general, \textsl{entropy measures the rate of creation of information}. Roughly, if the states of a system are described by iteration of a map, states that may be \textsl{indistinguishable} at some initial time may diverge into clearly \textsl{different} states as time passes. Entropy measures the \emph{rate} of creation of states. In the mathematical language it measures the \textsl{rate of divergence of orbits of a map}.\\

There are \emph{two} concepts of entropy for maps: \emph{topological} and \emph{measure-theoretic}. We shal consider mainly the first here due to Ghys, Langevin and Walczak (see \cite{ghys}). The idea is to \emph{define the entropy of the foliation as the entropy of its corresponding holonomy groupoid (or equivalently its holonomy pseudogroup)} in close analogy to the definition of the entropy of a map. The topological entropy of a foliation is closely related to two other notions: the \textsl{growth type} of its leaves (assuming that the leaves have Riemannian metrics we study the ``evolution'' of their volume) and the existence of \textsl{transverse holonomy invariant probabilistic measures}. In fact zero entropy implies the existence of a transverse invariant measure. We shall not discuss further these notions here.\\

Now suppose $f$ is a map of a compact manifold into itself. To measure the number of orbits one takes an empirical approach, not distinguishing $e$-close points for a given $e>0$. If $x$ and $y$ are two indistinguishable points, then their orbits $\{f^{k}(x)\}_{k=1}^{\infty }$ and $\{f^{k}(y)\}_{k=1}^{\infty }$ will be distinguishable provided that for some $k$, the points $f^{k}(x)$ and $f^{k}(y)$ are at distance grater than $e$. Then one counts the number of distinguishable orbit segments of length $n$ for fixed magnitude $e$ and looks at the growth rate of this function of $n$. Finally one improves the resolution arbitrarily well by letting $e\rightarrow 0$. The value obtained is called \emph{the entropy of $f$} and it measures the asymptotic growth rate of the number of orbits of finite length as the length goes to infinity.\\

More formally, let $(X,d)$ be a compact metric space and let $f:X\rightarrow X$ be a continuous map and $e$ a positive real number. We say that a set $E\subset X$ is \emph{$(n,e)$-separated by f} if whenever $x,y$ are two points of $E$, then $d(f^{k}(x),f^{k}(y))\geq e$ for some $0\geq k\geq n$. The maximum number of pairwise $e$-distinguishable orbits of length $n$ is
$$S(f,n,e):=sup\{card(E)|E is (n,e)-separated by f\}$$
\emph{This number is finite because $X$ is compact}. Its growth rate $h$ is by definition
$$h(f,e):=lim_{n\rightarrow \infty}sup\frac{1}{n}logS(f,n,e)$$
This number is possibly infinite and nondecreasing as $e\rightarrow 0$.\\
{\bf Definition:}The \emph{entropy of $f$} is the nonnegative or $\infty $ number
$$h(f):=lim_{e\rightarrow 0}h(f,e)$$

{\bf Example 1:} If $f$ is an \textsl{isometry} then its entropy is zero.\\

{\bf Example 2:} If $A$ is an $n\times n$ matrix with integer entries, then $A$ defines an endomorphism of the n-torus $T^{n}={\bf R}^{n}/{\bf Z}^{n}$. If $l_{1},...,l_{n}$ are the eigenvalues of $A$, then $h(f)=\sum _{|l_{i}|>1}log|l_{i}|$.\\

The next step is to define the \emph{entropy of transformation groups}. Let $G$ be a group of homeomorphisms of the metric space $(X,d)$ with finite generating set $G'$. We assume that this generating set is \textsl{symmetric}, namely $G'$ contains the identity transformation and is invariant under the operation of passing to the inverse. We set $G_{n}$ equal to the set of elements of $G$ that can be written as \emph{words of length $\leq n$ in elements of $G'$}. We say that points $x,y \in X$ are \emph{$(n,e)$-separated by $G'$} if there exists $g\in G_{n}$ such that $d(g(x),g(y))\geq e$. A subset $E$ of $X$ is \textsl{$(n,e)$-separated by $G'$} if each pair of distinct points in $E$ is so separated. The supremum of the cardinalities of such sets is denoted $S(G,G',n,e)$ and we define the \emph{entropy $h$} as
$$h(G,G'):=lim_{e\rightarrow 0}lim_{n\rightarrow \infty }sup\frac{1}{n}logS(G,G',n,e)$$

Then we can define the \emph{entropy of pseudogroups}. We fix a (possibly non-compact) metric space $(X,d)$ together with a pseudogroup $\Gamma $ of local homeomorphisms of $X$ that admits a finite symmetric generating subset $\Gamma '$ containig the identity. For each positive integer $n$, let $\Gamma _{n}$ denote the collection of elements of $\Gamma $ that can be obtained by composition of \textsl{at most} $n$ elements of $\Gamma '$ and let $\Gamma _{0}$ consist of the identity of $X$. Now let $e>0$. Points $x,y \in X$ are said to be $(n,e)$-separated by $\Gamma '$ if there exists $f\in \Gamma _{n}$ whose domain contains $x$ and $y$ and such that $d(f(x),f(y))\geq e$. A subset $E$ of $X$ is defined to be $(n,e)$-separated if every pair of distinct points $x,y$ in $E$ are $(n,e)$-separated. Let then $S(n,e)$ be the supremum of the cardinalities of the $(n,e)$-separated subsets of $X$. Then the \emph{entropy of $\Gamma $ with respect to $\Gamma '$} is by definition the nonnegative or $\infty $ \textsl{integer}
$$h(\Gamma ,\Gamma '):=lim_{e\rightarrow 0}lim_{n\rightarrow \infty }sup\frac{1}{n}logS(n,e)$$
One can prove that this number is \emph{independent} of metric $d$ compatible with the topology of $X$ for \textsl{regular pairs} $(\Gamma ,\Gamma ')$ (for the proof see \cite{CC} p353).\\

One then can define \emph{the entropy of a compact foliated manifold $($F$,$M$)$ with respect to a regular foliated atlas} $(U_{a}, \phi _{a})_{a\in A}$ as the entropy of its corresponding \textsl{total holonomy pseudogroup} $\Gamma _{U}$ (with respect to the regular foliated atlas). This is a \emph{finite} integer for $M$ compact. This definition can be generalised for foliated spaces.\\

{\bf Remark:}
Unfortunatelly (perhaps), the topological entropy of a foliation depends on the choice of a regular foliated atlas, hence \emph{it is not an invariant of the foliation}. That creates problems if one tries to relate that to the energy in physics in a similar fashion that the entropy of a statistical system is related to the energy. One then should apply a \emph{maximal entropy principle} in order to associate in a ``canonical'' way a number to a foliation; for example one may use the regular foliated atlas which \emph{maximises} the topological entropy (see \cite{Bayes}). It is \emph{merely the vanishing or non-vanishing} of the topological entropy which is an \textsl{invariant of the foliation(!)} and not the precise value of the topological entropy.\\

Let us give some examples: Cartesian products of compact manifolds considered as trivial foliations have zero topological entropy. The famous Reeb foliation of $S^{3}$ has zero topological entropy. Moreover if two foliations on the same manifold and of the same codimension are \textsl{integrable homotopic} and one of them has topological entropy zero, so has the other. Hence the condition ``zero topological entropy'' is an integrable homotopy invariant! (For the proof see \cite{CC} p362).\\

Perhaps the clearest example where one can see very clearly the role of the notion of $(n,e)$-separation and how it is used to define the topological entropy is the following elementary one from electromagnetism in dim-3 space (1-dim foliation): a \textsl{uniform} electric field given by the equation $\vec{E}=const$ is an example of a dynamical system with zero topological entropy, since the flow lines of the electric field are always ``parallel'' and they do not \emph{diverge,} namely the ``distance'' between them remains the same; hence there is no creation of information: states (i.e. flow lines which are the leaves of our 1-dim foliation) which are indistinguishable remain so for ever, using the notion of $(n,e)$-separation. This is in sharp contrast to the case of the flow lines of the electric field created by a \emph{point charge}. In this case the entropy is \emph{strictly positive} since there is creation of information: because the flow lines diverge, states (i.e. flow lines) which are very close to each other at some initial time and hence they are indistinguishable, they will eventually become distinguishable because the ``distance between them'' will increase. In the above example the flow lines in both cases are straight lines. Note that for the topological entropy what is important is not the shape of the flow lines (they could be arbitrary curves) but whether they diverge or not (namely if the distance between them remains the same or not). So an electric field which is not constant can still have zero entropy as long as the flow lines (being arbitrary curves) do not diverge. On the other hand the point charge creates an electric field which has non-zero topological entropy although its flow lines are straight lines.\\

There is a notion of \emph{geometric entropy} for a foliation using a \emph{leafwise Riemannian metric} and in fact the main theorem proved in \cite{ghys} indicates the relation between topological and geometric entropy for foliations. We shall not consider this here.\\

We now pass on to the main subject of this section, namely the relation between  topological entropy and the GV-class:\\

{\bf Proposition 2:}\\
If the compact foliated space $(M,F)$ has a \emph{resilient} leaf, then $F$ has \emph{positive} entropy.\\

For the proof see \cite{CC} p379.\\

The notion of resilient leaves can be generalised for the case of foliated spaces of arbitrary codim as follows: a loop $s$ on a leaf $L$ based at $x\in L$ is contracting if a holonomy transformation $h_{s}:D\rightarrow D$ associated to $s$ and defined on a suitable compact transverese metric disc $D$ centered at $x$, is such that $\cap ^{\infty}_{n=1}h^{n}_{s}(D)=\{x\}$. The leaf $L$ is \emph{resilient} if it has such a contracting loop and $L\cap (intD-\{x\})\neq \emptyset .$\\

Combining this with Duminy's theorem (for \emph{codim-1 case}) we get the following:\\

{\bf Corollary:}\\
If $(M,F)$ is a compact ($C^2$-)foliated manifold of \textsl{codim-1}, then \emph{zero entropy implies GV(F)=0}.\\

The \textsl{converse is not true}. A counterexample is the famous \emph{Hirsch foliation of the 3-torus}. This is a solid torus with a wormhole drilled out that winds around twice longitudinally while winding once meridionally, having a codim-1 foliation with the leaves being \textsl{``pair of pants''} 2-manifolds (see \cite{CC} p371). The Hirsch foliation has a resilient leaf, hence positive entropy. It is however transversely affine and hence its GV-class vanishes.\\

\section{Black Holes and Strings}

We would like to start this section by considering M-Theory first and address the following question: \emph{how much noncommutativity do we need} for M-Theory?\\

Following the original Connes-Douglas-Schwarz article, the answer will probably be: \emph{not very much}. In fact what these authors really considered was noncommutative \emph{deformations} of commutative algebras given by the Moyal bracket for instance. In their section 4.1 discussion where they considered the equivalent picture involving (codim-1) foliations of the 2-torus, what they used was in fact the moduli space of \emph{linear foliations} of the 2-torus. This is a \emph{2-manifold} and the GV-class (which would have been a \emph{3-form} for a codim-1 foliation) vanishes, so it is of no use. Yet the notion of topological entropy is useful; \textsl{in fact the topological entropy of all these linear foliations of the 2-torus is zero}. Let us define linear foliations on the 2-torus first and then explain why they have zero entropy: a constant vector field $\tilde{X}\equiv (a,b)$ on ${\bf R^2}$ is invariant by all translations in ${\bf R^2}$, hence passes to a well-defined vector field $X$ on $T^{2}={\bf R^2}/{\bf Z^2}$. We assume $a\neq 0$. \textsl{The foliation $\tilde{F}$ on ${\bf R^2}$ produced by $\tilde{X}$ has as leaves the} \emph{parallel} \textsl{lines of slope $b/a$}. (Just recall the usual topological description of the 2-torus as a rectangle with opposite sides identified). This foliation is also invariant under translations and passes to the foliation $F$ on $T^2$ produced by $X$. Since these leaves are \textsl{parallel} in the usual sense lines, there is no creation of information, hence the topological entropy is \emph{zero}. One can then consider two cases, rational and irrational slope (in both cases the entropy is zero): If the slope is rational, these foliations are in fact fibre bundles over $S^1$. If the slope is irrational, then each leaf is a 1:1 immersion of ${\bf R}$ and is everywhere dense in $T^2$ (Kronecker's theorem).\\

This is in accordance to the equivalent 11-dim supergravity picture presented in \cite{CDS} section 6, where it was argued that \emph{these} compactifications on  the noncommutative 2-torus (namely linear foliations of the 2-torus) correspond to turning on a \emph{constant} background field $C$ which is the 3-form potential of 11-dim supergravity. The 3-form field $C$ is constant because we require existence of BPS states preserving \emph{maximal} supersymmetry. (Turning on a background field $C$ which is an arbitrary 3-form would correspond mathematically to arbitrary foliations and physically to the case where not all of supersymmetries are preserved). Now if $C$ is constant then its field strength $dC$ of course is zero. That is \textsl{strongly reminiscent} of \textsl{foliations defined by closed forms} and in this case one is very tempted to recall all our previous discussion concerning \emph{codim-1 foliations}, Duminy's theorem and foliations with no holonomy. This is certainly not a proof, we just try to draw some rather striking analogies, keeping in mind that the situation for foliations of codim grater than one may be different; well, unfortunately rather few things are known for foliations of codim grater than 1.\\

Now we would like to make a connection with physics of the above discussion.
We have learned from \cite{CDS} that \textsl{M-Theory}
\emph{admits compactifications on noncommutative spaces}, tori more
specificaly. M-Theory on
the other hand is a more general theory than superstring theory, namely all 5
known consistent superstring theories can be derived from M-Theory and they are related by various string dualities. In a number of articles in the past, see
for example \cite{Hor1}, \cite{Hor2}, \cite{Hor3} and \cite{Hor4}, string theory claims it can give an answer to the origin of the black hole entropy formula due to Beckenstein-Hawking (see \cite{Town} and references therein).\\

Starting with a brief review, there is a striking similarity between the laws of black hole (abreviated to ``BH'' in the sequel) mechanics
$$dM=\frac{1}{8\pi G}\kappa dA$$
where $\kappa $ is surface gravity, $A$ is the area of the event horizon,
$G$ is Newton's constant
and the laws of thermodynamics
$$dE=TdS$$
Hawking in the 70's showed that this was more than an analogy with his
discovery that BH's indeed radiate a thermal spectrum with (Hawking) temperature
$$T_{H}=\frac{\hbar \kappa}{2k_{B}\pi }$$
where $k_{B}$ is Boltzman's constant and $\hbar $ is Planck's constant.
This implies that BH's have an entropy
$$S_{BH}=\frac{A}{4G\hbar}$$
But we know that the laws of thermodynamics are an approximation of the more fundamental laws of statistical physics. Hence we would like to have a description based on statistical mechanics, namely we want to find the
$$e^{S_{BH}}$$
quantum states associated with a BH.\\

Superstring theory claims it can give an answer for the origin of these
states, at least in some cases, based on \emph{S-duality.}\\

The argument briefly goes as follows:
Strings are 1-dim objects and their fluctuations produce all (in fact much
more than the) known particles. When the string is quantized in \emph{flat} spacetime, one finds an $\infty $ tower of states.
These include a \emph{finite} number of massless fields which include a \emph{scalar} field (``dilaton'') whose asymptotic value determines the string coupling $g$, \emph{gauge fields} and a spin-2 field identified with the graviton. Thus one mode of the string corresponds to a \emph{linearised} perturbation of the metric.
$\forall N$ integer there are \emph{massive} states with
$$M^{2}=\frac{N}{l_{s}^{2}}$$
where $l_{s}$ is the \textsl{string scale}, the \emph{``length''} of the string
(of the order of Planck's length).
These massive states have a large degeneracy.
The number of states with mass $M$ grows like $e^{M}$.
This was in fact the first clue that an explanation for the BH entropy might
come from superstrings.

But the number of BH states should grow like $e^{M^2}$. The rest of them will
come from the \emph{non-perturbative} sector of string theory, using results
from \textsl{string dualities.}

It was known for some time that string theory has also \emph{soliton}
solutions.
Solitons are static, finite energy classical solutions to the field equations. In string theory one has many types of solitons, including \emph{BH's} and
\emph{monopoles.}

To simplify our discussion we shall assume \emph{no backcreation}, namely
the mass ($=$total energy) of the BH is \textsl{constant}. This is true for
\emph{``extremely charged''} BH's and we shall focus our attention on them.

This assumption makes it reasonable to count only \emph{BPS states} in
superstring theory, since these have the important property that their mass
cannot receive quantum corrections.

Now the next problem which appears is the following: due to the coupling
to the dilaton, the extremal limit of BH solutions coming from superstring
theory with a \textsl{single charge} are quite different from the
well-known Reissner-Nordstrom solution from GR; precisely the problem is
that the event horizon becomes \emph{singular}.

The proposed way out then was to assume {\bf multi-charged} BH's. This
stabilizes the dilaton and it remains finite on the horizon in the extremal
limit. Hence the event horizon remains non-singular with finite area.
In this case then one can happily apply the usual quantization procedure for
solitons used in field theory, namely quantize only their zero modes.
But how many charges (i.e. gauge fields) does one have in string theory?

In general there are 2 types of gauge fields: \emph{$NS$} and \emph{$R$}.
They both may
carry either \textsl{electric} or \textsl{magnetic} type of charge. They
differ in the way the dilaton couples to them. Many of the $NS$ charges come
from the compactified directions of the original 10-dim metric (in a way
similar to
the way the electromagnetic potential arises from the 5-dim metric in
Kaluza-Klein theories).

Let us now consider the masses of the solitons carrying 1 unit (in units where
$l_{s}=1$) of these charges. Obviously one has 4 possibilities (recall that
$g$ is the string coupling):

\begin{center}
\begin{tabular}{|c|c|}
\hline Charge $Q$ & Mass $M$ \\
(1 unit)         &      (in units where $l_{s}=1$)         \\ \hline
NS ``electric''         &     1        \\
NS ``magnetic''         &     $1/g^{2}$       \\
R  ``electric''         &     $1/g$       \\
R  ``magnetic''         &     $1/g$       \\  \hline
\end{tabular}
\end{center}

We shall make use of the important relation between Newton's constant and
string coupling constant
$$G\sim g^{2}l_{s}^{2}$$
So when the string coupling constant is \emph{small}, $NS$ magnetic and both
$R$ charges are very \emph{massive}. One then may ask:
What is the gravitational field of these objects at weak coupling
$g\rightarrow 0$?
This is determined by term $GM$ and the key observation is that
since $G\sim g^{2}$, then $GM\rightarrow 0$ as $g\rightarrow 0$
$\forall M<1/g^{2}$.
So spacetime associated to these solitons becomes {\bf flat}
in the weak coupling limit for the $NS$ electric and both $R$ charges.

It turns out that for \textsl{$NS$ electric} solitons, the weak coupling
description comes from
some of the perturbative string states already mentioned. The way that an
\emph{extremely charged BH} can be identified with a
\emph{perturbative string state in flat spacetime} comes
from an \emph{S-duality} argument: as one \textsl{increases}
the string coupling, the mass does not change (since $M$ is independent of $g$
classically and supersymmetry
forbids any quantum corrections). But the \emph{gravitational field}
the mass produces becomes much stronger and it is described by a
{\bf curved} spacetime with large curvature.

Now we turn to \emph{$R$ charges}. What is the flat spacetime description of
a BH with \emph{$R$ charge?}
These cannot be perturbative string states since strings do not couple to
$R$ charges but to their field strengths.
In fact they turn out to be the \emph{$D$-branes}.\\

In practise then one usually starts with a BPS state with multiple $R$ charges, denoting them
$Q_{i}$ at \textsl{weak coupling}, which, as we explained above,
corresponds to \emph{flat} spacetime. This can be seen as a bound state of
several $D$-branes. As we increase the string coupling,
the gravitational field becomes stronger and the metric becomes an extremal BH
with non-singular horizon of finite area.

How many BPS states does one then have in weakly coupled string theory with
charges $Q_{i}$?

Let us consider the particular example calculated in \cite{Hor1}, namely a
\emph{5-dim black hole} with \emph{3 charges}. To understand the quantum states associated
with this black hole, it is more convenient to consider a 6-dim black string and the entropy of this black string will be the same as the black hole obtained
by dimensional reduction. Since string theory requires 10-dims, we assume that
the remaining 4 dimensions are compactified on a fixed torus of volume $(2\pi )^{4}V$ which is constant. We also assume that the dilaton is constant. The only non-trivial fields are the metric and the 3-form $H$. What happens in the compactification is the following: in addition to the usual Kaluza-Klein gauge field which arises from the metric, there is another gauge field coming from the dimensional reduction of the 3-form $H$ and that appears as a massless field. From the way that a string couples to the metric and $H$, one can show that a string with momentum in a compact direction carries \textsl{electric} Kaluza-Klein charge, while a string that winds around a compact direction carries \textsl{electric} charge associated with $H$. For 6-dim solutions with a spacelike translational symmetry, $H$ can carry both electric and magnetic charges which are proportional to the integral $\int *H$ and $\int H$ over the asymptotic 3-sphere
 in the  space orthogonal to the symmetry direction. These charges are quantized and take integer values which we shall denote $Q_1$ and $Q_5$. So we consider
the electric and magnetic charges associated to the massless field coming from the dimensional reduction of the 3-form $H$, we ignore the Kaluza-Klein field coming from the dimensional reduction of the metric and the 3rd charge will be
obtained by adding momentum along the string later. The solution to the low energy string equations turns out to be
$$ds^{2}=-(1-\frac{r_{+}^{2}}{r^{2}})dt^{2}+(1-\frac{r_{-}^{2}}{r^{2}})dx^{2}+(1-\frac{r_{+}^{2}}{r^{2}})^{-1}(1-\frac{r_{-}^{2}}{r^{2}})^{-1}dr^{2}+r^{2}d\Omega _{3}^{2}$$
This metric is similar to the 5-dim Reissner-Nordstrom solution. There is an event horizon at $r=r_{+}$ and an inner horizon at $r=r_{-}$. It is static, spherically symmetric and translationally invariant along the $x$-direction. The parameters $r_{+}$ and $r_{-}$ are related to the charges by
$$Q_{1}Q_{5}=\frac{r_{+}^{2}r_{-}^{2}V}{g^{2}}$$
The extremal limit corresponds to $r_{+}=r_{-}\equiv r_{0}$. If we periodically identify $x$ with period $2\pi R$, the extremal ADM energy is
$$E_{0}=\frac{2r_{0}^{2}RV}{g^{2}}$$
where we have used the fact that the 6-dim Newton's constant is
$$G=\frac{\pi ^{2}g^{2}}{2V}$$
in units with $l_{s}=1$, where $l_{s}$ is the string length. In this case the event horizon has \emph{zero area}. However, unlike the extreme solutions with a single charge the curvature does not diverge at the horizon in the extremal limit. The horizon area vanishes simply because the length in the $x$-direction shrinks to zero. \textsl{To obtain an extremal solution with non-zero area, we can add momentum along the string and this} \emph{provides the 3rd charge} upon to dimensional reduction to 5-dims. Since the extremal solution is boost invariant, we cannot add momentum by boosting it. Instead we start with the nonextremal solution, apply a boost $t=\tilde{t}cosh \sigma +\tilde{x}sinh \sigma $, $x=\tilde{x}cosh \sigma +\tilde{t}sinh \sigma $ and identify $\tilde{x}$ with period $2\pi R$. The ADM energy of these solutions is
$$E=\frac{RV}{2g^{2}}[2(r_{+}^{2}+r_{-}^{2})+cosh 2\sigma (r_{+}^{2}-r_{-}^{2})]$$
and the $x$ component of the ADM momentum is
$$P=\frac{RV}{2g^{2}}sinh 2\sigma (r_{+}^{2}-r_{-}^{2})$$
The horizon area is now
$$A=4\pi ^{3}r_{+}^{2}Rcosh \sigma \sqrt {r_{+}^{2}-r_{-}^{2}}$$
and the Hawking temperature is
$$T_{H}=\frac{\sqrt {r_{+}^{2}-r_{-}^{2}}}{2\pi r_{+}^{2}cosh \sigma}$$
The extremal limit is obtained by taking $r_{-}\rightarrow r_{+}$ keeping $P$
fixed, which requires $\sigma \rightarrow \infty $. The resulting solutions have energy
$$E_{ext}=E_{0}+P$$
Since the energy is increased by an amount equal to the momentum, the effect of boosting and taking the extremal limit is to add a null vector to the total energy. This can be viewed as adding \textsl{right moving} momentum only. In the extremal limit, the Beckenstein-Hawking entropy is
$$S_{BH}=\frac{A}{4G}=2\pi \sqrt {\frac{r_{0}^{4}VPR}{g^{2}}}=2\pi \sqrt {Q_{1}Q_{5}PR}$$
We now assume that \emph{$P$ is quantized}, $P=n/R$, since this will be the case for the quantum states we wish to count. So the entropy is simply
$$S_{BH}=2\pi \sqrt {Q_{1}Q_{5}n}$$
This expression depends \emph{only on the charges} and it's independent of both the volume of the compactified space and the string coupling. \textsl{The dimensional reduction of this extremal 6-dim black string with momentum gives a 5-dim extremal black hole with the same entropy}.\\

So far we have seen the properties of the classical solution. We shall now \emph{count states in string theory} and we shall try to show that it produces the \emph{same answer}.\\

The solution described arises in all string theories, since they all include an $NS$ field $H$. Yet the \emph{magnetic} $NS$ charge does not have a flat spacetime description and for that reason we shall consider the type IIB theory which has another 3-form which is an $R$ field. The solution for the black string is identical and we shall count its states in the weak coupling region. So we start from 10-dim flat spacetime and compactify 4-dims on a torus of volume $(2\pi )^{4}V$ and one direction on a circle of circumference $2\pi R$ which is much larger than the other four. It turns out that the objects which carry the charges $Q_{1}$ and $Q_{5}$ are respectively a $D$-string wrapping $Q_{1}$ times around the circle with radius $R$ and a $D$5-brane which wrapps $Q_{5}$ times around the 5-torus. With a little bit more effort we finally come to the conclusion that \textsl{for large $n$} the result coincides with the Beckenstein-Hawking entropy formula derived from black hole physics considerations above:
$$S=2\pi \sqrt {Q_{1}Q_{5}n}$$
So now our discussion starts with the following question:\\

\emph{What happens if the compactified 5-torus is} {\bf noncommutative}?\\

We know after the appearence of \cite{CDS} that noncommutative tori \textsl{are allowed} as M-Theory, and hence superstring theory, compactifications.\\

Before trying to answer this question, let us mention that this case seems
\emph{very similar} to the case of the explanation of the \emph{integrality} of the \textsl{Hall conductivity} $\sigma _{H}$ related to the \emph{drift current} in the {\bf quantum Hall effect}
by Bellissard, who observed that the \textsl{Brilluin zones} in momentum space form \textsl{noncommutative 2-tori} when the magnetic flux is \textsl{irrational} (see \cite{Connes} Chapter IV section 6 and references therein). We still have electric and magnetic charges here but arising from the Abelian field $H$ which is a real valued 3-form instead of the usual electromagnetic potential which is a real valued 1-form.\\

First let us say that we do not have a clear picture on what happens in the GR
framework. In any case, these extra dimensions do not seem to be related to GR since for it, one only has 4 (or 5 in our example) dimensions. It is the
string theory answer which might be affected by this assumption.

Anyway, the first thing to note is the following: assuming that the answer is still the product of the 3 charges, the most obvious difference lies on the charge $Q_{5}$ since the $D$5-brane wraps around the noncommutative 5-torus. Our suggestion is that this charge should be replaced by a \emph{topological invariant} for the \textsl{noncommutative 5-torus}. We know that foliations are classified up to homotopy by the Pontryagin classes of their corresponding normal bundle (this is Bott's theorem) but up to integrable homotopy they are classified by the Gelfand-Fuchs cohomology classes (or the invariant introduced in \cite{Z2}).\\

We think it is clear now that the foliation picture description of noncommutative spaces can at least give a feeling of what one should do. Remember that what
we are actually interested in is the number of some quantum states. One may expect that this should be related to the \emph{volume} of the 5-torus and this is still the same, no matter if we deform the algebra of functions in order to make it noncommutative. Yet if we deform it \textsl{``too much''}, namely the \emph{GV-class becomes non-zero}, then the topology becomes different, in fact we enter the region of noncommutative topology and that we believe \textsl{should affect the answer}. For the moment we cannot offer any more concrete argument. If the torus becomes noncommutative but the GV-class is still zero, then topologically we have no other topological charge available; we think of the non-vanishing of the GV-class as a \emph{``phase transition''} from the realm of \textsl{commutative} to the \textsl{noncommutative topology}. An additional element which we think supports this idea is also the fact that as we saw in the previous section, the non-vanishing of the GV-class is related to the appearence of topological entropy and it is not true that every noncommutative space, namely every foliation, has non-zero topological entropy.\\

An immediate problem is the following: in general the GV-invariant is a \emph{real} number and \emph{not an integer}. This is crucial since we are counting quantum states. So the GV-class should not be \textsl{directly related to the charge $Q_{5}$}, the relation must be more subbtle. From the work of Bellissard however on the quantum Hall effect we have learnt that one might use a \textsl{cyclic cocycle} on some noncommutative algebra which is \emph{integral}. This is the cyclic cocycle denoted $\tau _{2}$ in the relevant section in \cite{Connes} but this has no immediate relation with the GV-class, it is something purely algebraic (there is however the remark 12 in \cite{Connes} p366 which makes a connection with foliations with a leaf $L$ of non-positive curvature).\\

The last comment is the following: we saw that foliated manifolds may have non-zero topological entropy and commutative spaces have zero topological entropy. Naively one might think then that some quantum states should be used to give rise to the
(possibly) non-zero topological entropy of the compactified dimensions assumed to form a foliated manifold. We also know fron the work of A. Connes that the GV-class is related to the
von-Neumann algebra of the foliation and that is related to measure theory. For the moment the whole situation seems very intriguing but unfortunately we have no concrete answers.\\

\section{Appendix}

We shall try to exhibit a sketch for the proof of Duminy's theorem (which recall, applies to codim-1 foliations). We shall give some definitions also from the theory of \emph{levels} (or \emph{``architecture'') of foliations}. Our reference is \cite{CC}.\\

Let $(M,F)$ be a codim-1 foliation $F$ on a closed smooth oriented $n$-manifold $M$. A subset $X$ of $M$ is called \emph{F-saturated} if it is a union of leaves of $F$. It is called a \emph{minimal} set if it is closed, nonempty, $F$-saturated having itself no proper subset with these properties. (Example: in a fibre bundle each leaf, the leaves in this example are just the fibres, is a minimal set).\\

A leaf that belongs to a minimal set of $F$ is said to be at \emph{level 0}. The union of all leaves of level 0 is denoted $M_0$ which is compact. Now we set $U_{0}:=M-M_{0}$. In this case either $U_{0}=\emptyset $ or $F$ restricted to $U_{0}$ denoted $F|_{U_{0}}$ has at least one minimal set. In the first case $M=M_{0}$. In the second, let us denote $M_{1}$ the union of $M_{0}$ and all of the minimal sets of $F|_{U_{0}}$. $M_{1}$ is closed in $M$, hence it is also compact and we let $U_{1}:=M-M_{1}$. Again either $M=M_{1}$ or we obtain a compact $F$-saturated set $M_{2}$ as the union of $M_{1}$ and all minimal sets of $F|_{U_{1}}$. Then inductively one gets the following\\

{\bf Theorem:}\\
There is a unique filtration
$$\emptyset :=M_{-1}\subset M_{0}\subseteq M_{1} \subseteq ... \subseteq M_{k}\subseteq ... \subseteq M$$
of $M$ by compact $F$-saturated subsets such that:\\
1. $M_{k}-M_{k-1}$ is the union of all minimal sets of $F|_{(M-M_{k})}$ for all $k\geq 0$\\
2. If $M_{k}=M_{k+1}$ for some $k\geq 0$, then $M_{k}=M_{k+p}=M$ for all $p\geq0$\\

We define the leaves of  $F|_{(M-M_{k})}$ to be at \emph{level $k$}. Denote $M_{*}:=\cup _{k=0}^{\infty }M_{k}$ and we say that the leaves of $F|_{M_{*}}$ are at \emph{finite level}. The leaves (if any) in $M_{\infty }:=M-M_{*}$ are said to be at \emph{infinite level}. $M_{\infty }$ is either empty or it contains  uncountably many leaves. Moreover it has no interior, hence one cannot continue finding minimal sets at infinite levels and as a subset of $M$ it is a Borel set and Lebesgue measurable. For an $F$-saturated measurable set $X\subseteq M$, one can show that the GV-class $GV(F)$ can be \emph{intergrated} over $X$ to define a cohomology class $GV(X,F)\in H^{3}(M;{\bf R})$. This is NOT obvious since the GV-class GV(F) is a 3-form (for codim-1 foliations) and $X$ is a measurable subset of $M$. One uses some aspects of Poincare duality (see \cite{CC} p193). More cencretely then assuming $M$ to be closed, oriented smooth $n$-manifold, one version of Poincare duality identifies the vector spaces
$$H^{q}(M;{\bf R})=Hom_{\bf R}(H^{n-q}(M;{\bf R}),{\bf R})$$
as follows: given $[\phi ]\in H^{q}(M;{\bf R})$ represented by the $q$-form $\phi $, define
$$[\phi ]:H^{n-q}(M;{\bf R})\rightarrow {\bf R}$$
by
$$[\phi ]([\psi ]):=\int _{M}\phi \wedge \psi $$
and then define $GV(X,F):H^{n-3}(M;{\bf R})\rightarrow {\bf R}$ via
$$GV(X,F):=\int _{X}\eta \wedge d\eta \wedge \psi $$
One can then view the GV-class as an $H^{3}(M;{\bf R})$-valued countably additive measure on the $\sigma $-algebra of Lebesgue measurable $F$-saturated sets. This measure satisfies $GV(M,F)=GV(F)$, namely we get the original GV-class of the foliation if we apply the construction just described to the $F$-saturated set $M$ itself. This is used in Duminy's theorem to prove that $GV(F)$ is zero unless some leaf is resilient. One has the following level-decomposition sum:
$$GV(F)=GV(M_{\infty },F)+\sum _{k=0}^{\infty }GV(M_{k},F)$$
If no leaf is resilient, the minimal sets are either proper leaves or open without holonomy. This is used to prove that $GV(M_{k},F)=0$ for all $k\geq 0$, hence $GV(F)=GV(M_{\infty },F)$ and then the last step is to prove that this class vanishes. In fact Duminy proves that $GV(M_{\infty },F)=0$ for all $C^2$-foliations, whether or not they have resilient leaves.\\

The last comment is the following: it is a delicate issue in the theory of foliations the difference between functors $\Gamma _{q}(-)$ and $Fol_{q}(-)$ from spaces to topological categories. The former comes by dividing the set of all \emph{Haefliger $q$-cocycles} (which for the special case of a Haefliger structure being a foliation is essentially the \emph{germinal holonomy} or the \emph{holonomy groupoid} of the foliation) $H^{1}(-,\Gamma _{q})$ by \emph{(ordinary) homotopy} whereas for the second one has to divide by \emph{integrable homotopy}. The GV-class \textsl{in general} (there are some differences for codim-1 foliations and for 3-manifolds as we have already explained) is integrable homotopy invariant hence it naturally lives in $Fol_{q}$. Foliations up to homotopy are essentially classified by the Pontryagin classes of their normal bundle (this is a consequence of the Bott-Haefliger theorems mentioned in \cite{Z1}). We suspect then that the K-Theory constructed using the Quillen-Segal construction and the topological category $\Gamma _{q}(M)$ for a manifold $M$ might be closely related to Atiyah's ordinary K-Theory. What seems to be more interesting is probably the K-Theory arising from $Fol_{q}(M)$. This \emph{was not included} in our list of K-Theories in \cite{Z1}, although it appears that it is probably the most appropriate one since we were using the GV-class.\\

For an \emph{open} smooth $n$-manifold $M$ one has a way to determine $Fol_{q}(M)$ by means of $\Gamma _{q}(M)$ using the Philips, Gromov, Haefliger (PGH) theorem as follows (see \cite{Bott} for more details):\\

Let $F$ be a codim-$q$ foliation of $M$. Then the tangent bundle $TM$ of $M$ splits as $TM=F\oplus \nu (F)$ where $\nu (F)$ is the normal bundle of the foliation. Let $g_{M}:M\rightarrow BGL(n;{\bf R})$ denote the \emph{Gauss map} that determines the tangent bundle $TM$ of $M$. From the splitting of $TM$ we deduce that the Gauss map $g_{M}$ admits a homotopy lift $G_{M}:M\rightarrow BGL(q;{\bf R})\times BGL((n-q);{\bf R})$. Since $\nu (F)$ is the normal bundle of the foliation, then $G_{M}$ admits a second homotopy lift $\tilde{G_{M}}:M\rightarrow B\Gamma _{q}\times BGL((n-q);{\bf R})$. Then one has the following commutative diagram:

\begin{equation}
\begin{CD}
M@>\tilde{G_{M}}>>B\Gamma _{q} \times BGL((n-q);{\bf R})\\
@V\cong VV     @VVB\nu \times id V\\
M@>>G_{M}>BGL(q;{\bf R})\times BGL((n-q);{\bf R})\\
@V\cong VV     @VVpV\\
M@>>g_{M}>BGL(n;{\bf R})\\
\end{CD}
\end{equation}

Then the (PGH) theorem says that there is a 1:1 correspondence between elements of $Fol_{q}(M)$ and the set of homotopy classes of homotopy lifts $\tilde{G_{M}}$ of $g_{M}$.\\

The map $B\nu :B\Gamma _{q}\rightarrow BGL(q;{\bf R})$ is the corresponding continuous map on the classifying spaces of the functor $\nu :\Gamma _{q}\rightarrow GL(q;{\bf R})$ which defines the normal bundle of the foliation just by considering the Jacobian of any local diffeomorphism (restricted to each leaf this is just the infinitesimal germinal holonomy of the leaf as we described above). (See also \cite{Z1}).\\

Let us now pass on to the notion of \emph{tautness} for foliations of arbitrary codim ($>1$). There is exactly the same definition for geometric tautness using Riemannian metrics. Yet instead of topological tautness there is an analogous notion called \emph{homological tautness}, which reduces to the well-known definition for the codim-1 case (see \cite{CC} p266). Sullivan's theorem then states that a codim-$q$ foliation is geometrically taut iff it is homologically taut. An analogous characterisation using differentail forms also exists in this case, hence a codim-$q$ foliation $F$ on a closed smooth $n$-manifold $M$ is taut iff there exists an $(n-q)$-form $\theta $ on $M$ which is \emph{$F$-closed} and transverse to the foliation. Transverse means it is non-singular when restricted to all leaves and the condition of being $F$-closed is a weaker condition than being closed; it means that $d\theta =0$ whenever \textsl{at least $(n-q)$} (i.e. same number as dim of the foliation) \textsl{of the vectors are tangent to $F$}.\\

\end{document}